\documentclass[prd,letterpaper,tightenlines,showkeys,showpacs,preprintnumbers,nofootinbib,floatfix,psfig]{revtex4}

\usepackage{amsmath,amssymb,amsfonts,graphicx,pifont,dcolumn}
\usepackage{epstopdf}
\usepackage{color}
\usepackage{bm}
\usepackage{cancel}
\usepackage{ulem}
\usepackage{comment}
\RequirePackage{slashed}

\begin{document}

\title{Pure glueball states  in a Light-Front holographic approach}

\author {Matteo Rinaldi}

\affiliation{Dipartimento di Fisica e Geologia. Universit\`a degli 
Studi di Perugia and INFN section of Perugia. Perugia Via A. Pascoli, 
I-06123,Italy}

\author{ Vicente Vento }

\affiliation{Departamento de F\'{\i}sica Te\'orica-IFIC, Universidad de
Valencia- CSIC, 46100 Burjassot (Valencia), Spain.}

\date{\today}

\begin{abstract}

A phenomenological analysis of the scalar glueball and scalar meson 
spectra is carried out by using the AdS/QCD framework in the bottom-up 
approach. The resulting spectra are in good agreement for glueballs with lattice QCD results and for mesons with PDG data. We make use of the relation between the mode functions in 
AdS/QCD
and the wave functions in Light-Front $QCD$ to discuss the mixing of
glueballs and mesons. The results of our investigation point out that above 2 GeV scalar particles will appear in almost degenerate pairs of unmixed glueball and mesons states 
leading to an interesting phenomenology whereby gluon dynamics could be well investigated.

\end{abstract}



\maketitle

\section{Introduction}

Glueballs have been a matter of theoretical study and experimental 
search 
since the formulation of the theory of the strong interaction
Quantum Chromodynamics
(QCD)~\cite{Fritzsch:1973pi,Fritzsch:1975wn}, QCD sum
rules ~\cite{Shifman:1978bx,Mathieu:2008me}, QCD based models~\cite{Mathieu:2008me} 
and Lattice QCD computations both with sea
quarks~\cite{Gregory:2012hu} and in the pure glue theory~\cite{Morningstar:1999rf,Chen:2005mg,Lucini:2004my} have been used to
determine their spectra and properties. However, due to the lack of
phenomenological support  
much debate has been  associated with their properties~\cite{Mathieu:2008me}. Glueballs, if they exist,  will mix with meson
states of the
same quantum numbers, and therefore their direct characterization is
difficult 
to disclose. 

A  fruitful strategy to investigate 
non perturbative features of glueballs  is the use of QCD models inspired 
by the holographic conjecture. Recently we have used these so called 
AdS/QCD models to study the glueball spectrum~\cite{Vento:2017ice,Rinaldi:2017wdn}.
The holographic principle relies in a correspondence between a five dimensional 
classical theory with an AdS metric and a supersymmetric conformal quantum 
field theory with $N_C  \rightarrow \infty$. Since the latter is different from 
QCD, an approach called bottom-up is usually 
adopted~\cite{Polchinski:2000uf,Brodsky:2003px,DaRold:2005mxj,Karch:2006pv}. 
It consists  in properly 
modifying the five dimensional 
classical theory to  resemble  QCD as much as possible.
The different models constructed in this way differ in the 
strategy to break conformal invariance. Since the mesons masses are 
$\mathcal{O}(N^0_c)$ these models reproduce the essential features of 
the meson spectrum~\cite{Erlich:2005qh,deTeramond:2005su,Colangelo:2008us}\footnote{Large $N_C$ behavior of observables can be found in the lectures by J.L.~Goity \cite{Goity:2016ty} and references therein.}. Two of the most widely used models are the hard-wall 
 (HW) and the soft-wall (SW) models. The former  consists in introducing an explicit cut 
off in the fifth coordinate z,   $0 \leq z \leq z_{max} = 1/\Lambda_{QCD}$.  However, the 
HW model is not able to reproduce the 
Regge trajectories of the mesonic spectrum. To this aim, in the SW model,  a dilaton field  is introduced to softly 
break 
conformal invariance. Since the glueball masses are also $\mathcal{O}(N_C^0)$ these models were extended to study the glueball spectrum~\cite{Rinaldi:2017wdn,Colangelo:2007pt,Capossoli:2015ywa}.  Keeping in mind the relation between AdS/QCD models and the $1/N_C$ expansion of QCD other observables of $\mathcal{O}(N_C^0)$ have been investigated. In fact, since in the  the planar limit ($N_C\rightarrow \infty$) hadrons are stable and non-interacting, these AdS/QCD models are appropriate to evaluate structure functions and spectra. 
For example form factors (ffs), parton distribution functions (PDFs), generalized parton distribution functions and transverse dependents PDFs have been calculated with reasonable agreement with the data~\cite{Brodsky:2006uqa,Abidin:2009hr,Chakrabarti:2013dda,Rinaldi:2017roc,Bacchetta:2017vzh,deTeramond:2018ecg}. All these analyses were inspired by the investigation
 on hard deep inelastic scattering at small $x$  within AdS/QCD~\cite{Polchinski:2002jw}.
In addition, $\mathcal{O}(1/N_C^2)$ observables like the pomeron contribution to proton-proton scattering~\cite{Domokos:2009hm,Shuryak:2013sra}
 and the elastic cross section~\cite{Watanabe:2018owy} have been also described with holographic inspired models.

Following our investigations on the scalar glueball spectrum, we discuss in what follows some phenomenological consequences by comparing theoretical results with data. The experimentally
determined scalar "mesons" with spin parities $J^{PC}= 0^{++}$  
are known as the $f_0$ mesons~\cite{Patrignani:2016xqp}. Our aim is to 
describe the glueball lattice spectrum 
by means of the AdS/QCD correspondence and to compare it with the 
spectrum of 
$f_0$'s. Only when the masses of the glueballs and the $f_0$'s 
are close mixing is to be expected~\cite{Vento:2004xx}. However, 
if the masses are close, but
the dynamics generating the resonance states is different, mixing 
will not happen~\cite{Vento:2015yja}. 
Therefore, we are looking for meson and glueball states with similar 
masses but generated by different dynamics.  
These scalar particles will appear in the phenomenological spectrum as
mostly glueball or mostly meson.

We use for our analysis the bottom-up approach of the 
AdS/QCD 
correspondence~\cite{Polchinski:2000uf,Brodsky:2003px,DaRold:2005mxj}.
This approach describes  glueball and meson dynamics  in a very 
transparent 
fashion~\cite{Erlich:2005qh,Karch:2006pv,Rinaldi:2017wdn,
Colangelo:2007pt}. 
In section II we describe the scalar glueball spectrum  
and the meson spectrum in several AdS/QCD approaches. In section III 
we 
will discuss glueball-meson mixing and finally in the conclusions
we extract some consequences of our analysis.

\section{Scalar glueball and scalar meson spectrum in a bottom-up 
approach}

The  botton-up approach on the AdS/CFT correspondence starts
from QCD and  attempts to construct its 
five-dimensional holographic dual. One implements  duality in nearly 
conformal 
conditions defining QCD on the four dimensional  boundary and 
introducing a bulk 
space which is a slice of $AdS_5$ whose size is  related to 
$\Lambda_{QCD}$ 
~\cite{Polchinski:2000uf,Brodsky:2003px,Erlich:2005qh,DaRold:2005mxj}. 

\begin{figure}[htb]
\begin{center}
\includegraphics[scale= 0.27]{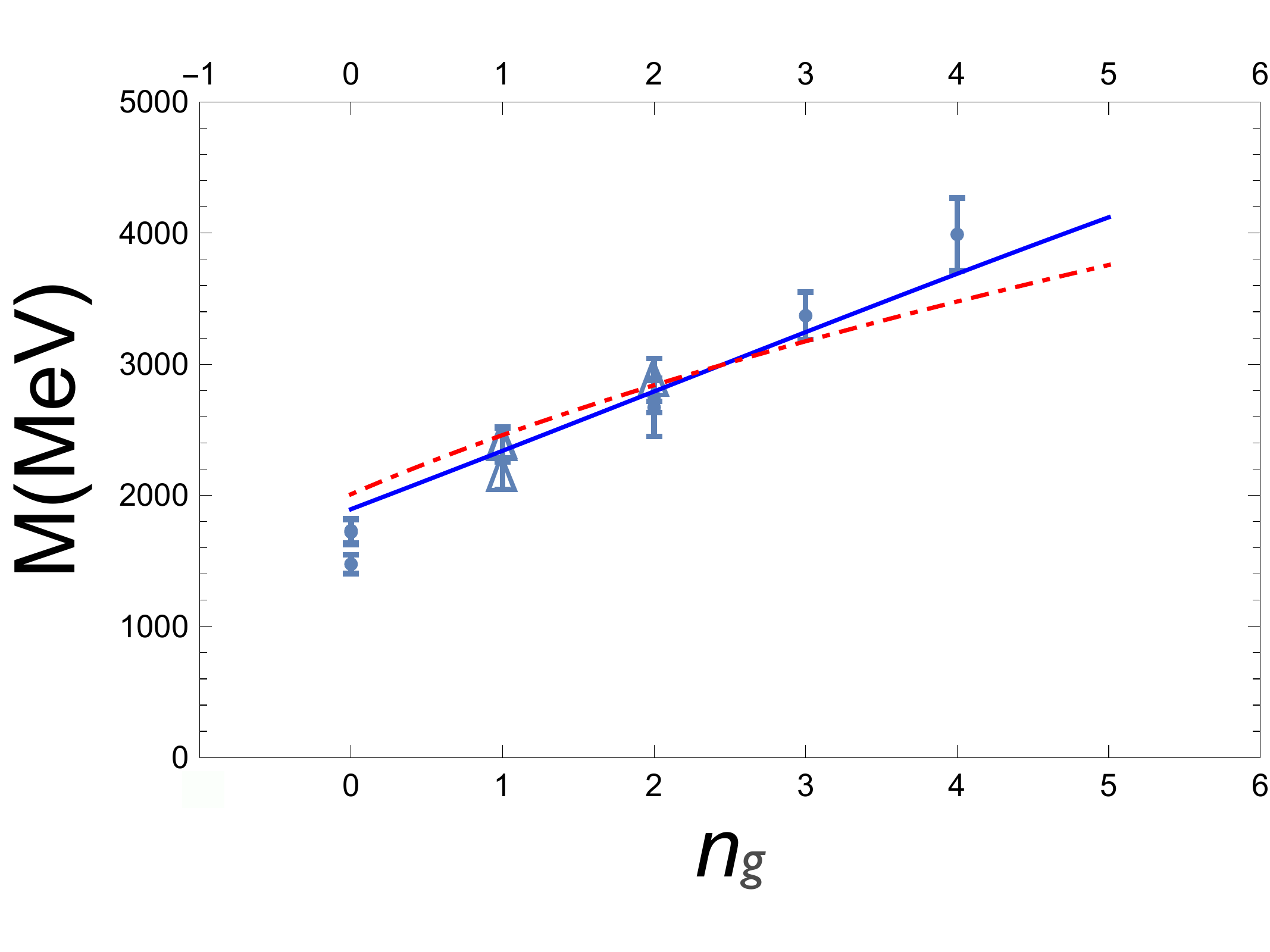} 
\end{center}
\caption{Glueball spectrum obtained with the soft wall model (dashed-dotted)~\cite{Colangelo:2007pt} 
and soft wall graviton model (solid)~\cite{Rinaldi:2017wdn}. These calculations were reported in ref. 
~\cite{Rinaldi:2017wdn}. 
The lattice data are from refs. 
~\cite{Morningstar:1999rf,Chen:2005mg,Lucini:2004my} as shown in 
Table~\ref{Gmasses}. The dots label the scalar glueballs and the 
triangles the tensor glueballs.} 
\label{GlueballFit}
\end{figure}

The metric of this space can be written as

\begin{equation}
ds^2=\frac{R^2}{z^2} (dz^2 + \eta_{\mu \nu} dx^\mu dx^\nu) + 
R^2 d\Omega_5,
\label{metric5}
\end{equation}
where $\eta_{\mu \nu}$ is the Minkowski metric and the size of 
the slice in the
holographic coordinate $0< z <  z_{max}$ is related to the scale of QCD,
$ z_{max} =\frac{1}{\Lambda_{QCD}}$. This is 
the so called hard wall approximation. Later on, in order to 
reproduce the Regge 
trajectories, the  so called soft wall approximation  was introduced 
~\cite{Karch:2006pv,Capossoli:2015ywa}. Within the bottom-up
strategy and in
 the soft wall approach, glueballs 
arising from the correspondence of  fields in $AdS_5$ have  been
studied in ref.
~\cite{Colangelo:2007pt}.  In our recent work we have 
described the scalar glueball spectrum as that of a graviton in $AdS_5$ with a 
modified soft wall metric~\cite{Rinaldi:2017wdn}.  In Fig.~\ref{GlueballFit} we 
reproduce the 
results of these references for the glueball spectrum.

In order to clarify the discussion of the present investigation let us recall the formalism to describe glueballs and mesons 
in  AdS/QCD soft wall models.

\subsection{The glueball in a soft wall model}
 For a scalar glueball, 
the dilaton the action is:

\begin{align}
 S = \int d^5 x~\sqrt{-g} e^{-\phi(z)} \Big[ g^{MN} \partial_M G~ 
\partial_N G+M^2_{5g} G^2   \Big]~,
\label{glact}
\end{align}
where here $g_{MN}$ is the metric of eq.~(\ref{metric5}), $G$ the 
scalar glueball field, $M^2_{5g}$  is the $AdS_5$ mass and the 
dilaton function is $\phi(z)=z^2$. In the case of scalars, 
the boundary operator conformal 
dimension $\Delta$, is related to $M_5$ as follows~\cite{Capossoli:2015ywa}:

\begin{align}
 \Delta = 2 \pm \sqrt{4 + R^2 M_5^2}~.
\end{align}
In the case of the scalar glueball, $\Delta = 4$ thus $M_{5g}=0$. The 
equation of motion from eq.~(\ref{glact}) for the glueball field reads:

\begin{align}
 \partial_M \Big[ \sqrt{-g} e^{-\phi(z)}g^{MN} \partial_N G   \Big]=0~.
\end{align}

The above equation can be rewritten  as a Schr\"odinger like equation:

\begin{align}
 -\Psi{''}(z)+ \Big[ z^2+ \frac{15}{4 z^2}+2  \Big]\Psi(z) = \mu^2 
\Psi(z)~,
\label{gsch}
\end{align}
where here 
\begin{align}
 G(z,x) = e^{i q \cdot x} \Psi(z) \left( \frac{z}{R} 
\right)^{3/2}e^{z^2/2}~.
\end{align}
In the above relation, $x$ is a four vector in the Minkowski space and 
 $-q^2 = \mu^2$ with $\mu$ an adimensional glueball mass. As can be seen in ref.~\cite{Colangelo:2007pt} the solution of eq.~(\ref{gsch})  leads to the following mode spectrum

\begin{align}
 \mu^2_k = 4k+8,
 \label{glueballsp}
 \end{align}
where $k = 0, 1, 2, \ldots$.  The corresponding normalized mode function for mode number  $k = n_g$ is
can be written as:

\begin{equation}
\Psi_{n_g}(z)=\sqrt{(n_g+1)(n_g+2)/2} \; e^{-z^2/2} \;  z^{5/2} \;
_1F_1(-n_g, 3, z^2)~.
\end{equation}

\subsection{The scalar meson in the soft wall model}
For a scalar meson, the 
dilaton the action is:

\begin{align}
 S = \int d^5 x~\sqrt{-g} e^{-\phi(z)} \Big[ g^{MN} \partial_M S~ 
\partial_N S+M^2_{5m} S^2   \Big]~,
\label{meact}
\end{align}
where  $S$ the 
scalar meson field and $M^2_{5m}$  is the $AdS_5$ mass. The conformal dimension for the scalar meson 
field is $\Delta = 3$ thus $M^2_{5g} R^2=-3$. The 
equation of motion from eq.~(\ref{meact}) for the scalar meson field 
reads~\cite{Colangelo:2008us},:

\begin{align}
 -\Psi{''}(z)+ \Big[ z^2+ \frac{3}{4 z^2}+2  \Big]\Psi(z) = \mu^2 
\Psi(z)~,
\end{align}
where here 
\begin{align}
 S(z,x) = e^{i q \cdot x} \Psi(z) \left( \frac{z}{R} 
\right)^{3/2}e^{z^2/2}~.
\end{align}
In the above relation 
 $-q^2 = \mu^2$ with $\mu$ an adimensional scalar meson mass. The solution of this equations~\cite{Colangelo:2008us} leads to the following mode spectrum,
by

\begin{align}
 m^2_k = 4k+6,
 \label{mesonsp}
\end{align}
where $k = 0, 1, 2, \ldots$. The  normalized mode function for mode number $k = n_m$ is

\begin{equation}
\Psi_{n_m}(z)=  \sqrt{2(n_m+1)} \; e^{-z^2/2} \;z^{3/2} \; 
_1F_1(-n_m, 2, z^2)~.
\end{equation}

\subsection{The soft wall graviton model}
In ref.~\cite{Rinaldi:2017wdn}, we discussed the possibility that the 
glueball field is dual to a graviton. However, in order to 
recover the above results for the scalar meson, it is 
convenient to generalize the background metric to

\begin{align}
\bar g_{MN} = e^{-\alpha^2 z^2/R^2} g_{MN}~,
\end{align}
with $\alpha^2 < 0$ in order to have bound states. The 
equation of motion is obtained by solving the Einstein equation for the 
above metric~\cite{Rinaldi:2017wdn}. Setting $R=1$ this equation reads

\begin{equation}
\frac{d^2 \phi}{d z^2}  + \left(\alpha^2 z - \frac{3}{z}\right) \frac{d \phi}{d 
z} +
\left(\frac{8}{z^2} + 6 \alpha^2 + \mu^2 + 4 \alpha^2 z^2\right) \phi - 
\frac{8}{z^2} e^{-\alpha^2 z^2} \phi =0.
\end{equation}
In this case one can also obtain a Schr\"odinger 
like equation by performing the change of function

\begin{equation}
\phi(z) = e^{\alpha^2 z^2/4}\left( \frac{z^2}{\alpha^2}\right)^\frac{3}{2} \Psi(z)
\end{equation}
which using the adimensional variable $t = a z /\sqrt{2}$ where $a = i \alpha$ becomes

\begin{align}
 -\Psi''(t)+\Big[\frac{8 e^{2t^2}}{t^2}-\frac{17}{4 t^2}+14-15 t^2  
\Big]\Psi(t)= \frac{2 \mu^2}{a^2} \Psi(t)~,
\end{align}
The mode spectrum has no analytical solution and was obtained numerically ~\cite{Rinaldi:2017wdn}.
We take $a=\sqrt{2}, t=z$ and the equation resembles those of the other approaches with $\mu$ 
an adimensional glueball mass.

\subsection{Phenomenological analysis}

As we have just recalled the AdS/QCD models provide us with a succession of mass modes 
of differential equations, which in general, one has to obtain 
numerically. In the case of glueballs an exception is  the standard soft wall model where the 
expression turns out to be analytic,  $\mu{^2}(n_g) = 4 n_g +8$, where $n_g$
is the corresponding mode number~\cite{Colangelo:2007pt}. In order to 
reach the experimental results we have to multiply the adimensional modes  $\mu(n_g)$ by an energy scale 
 $\varepsilon$, i.e.  $m^2(n_g) = \varepsilon^2 \mu^2(n_g) $
To determine $\varepsilon$ we  use  here the technique,  developed in
ref.~\cite{Rinaldi:2017wdn} . We display the spectrum obtained by 
fitting one  $AdS$  mode to a physical state,  for example in the case 
of glueballs we fit the lowest mode to the lowest lying glueball and 
determine an initial value for  $\varepsilon$. We then proceed by 
seeding the rest of the lattice data on the plot and by varying slightly 
$\varepsilon$ we get  a best fit to the whole spectrum. 
 The lattice data used are shown in Table~\ref{Gmasses} 
~\cite{Morningstar:1999rf,Chen:2005mg,Lucini:2004my}   
\footnote{We have not included the lattice results from the unquenched 
calculation 
~\cite{Gregory:2012hu} to be consistent, which however,  in this
range of
masses and for these quantum numbers are in agreement with the
shown results 
within errors.}.
We also use for the fit  the results for the tensor 
glueball states since the theory predicts degeneracy between the scalar 
an the tensor glueball for the soft wall models. In Fig.~\ref{GlueballFit} we show the results for all models analyzed.
The best fit for the soft wall model is obtained for $\varepsilon^\prime =710$ MeV.
The  soft wall graviton model  has no analytical mode solution, it is given by a numerical function  $m(n_g)= \varepsilon'' f(n_g))$, and the shown fit is for  $\varepsilon''=370$ MeV.
Both  models lead to a reasonable fit of the data. The difference
between them arises for heavy states. The soft wall dilaton model has a 
quadratic behavior that softens the slope at high energies. 
The soft wall graviton model has an almost linear behavior 
 showing no softening of the slope in the region analyzed.

\begin{table} [htb]
\begin{center}
\begin{tabular} {|c c c c c c c|}
\hline
$J^{PC}$& $0^{++}$&$2^{++}$&$0^{++}$&$2^{++}$&$0^{++}$&$0^{++}$\\
\hline
MP & $1730 \pm 94$ & $2400 \pm122$ & $2670 \pm 222 $&  & &  \\
\hline
YC & $1719 \pm 94$ & $2390 \pm124$ &  &  &  &  \\
\hline
LTW & $1475 \pm 72$ & $2150 \pm 104$ & $2755 \pm 124$& $2880 \pm 
164 $& $3370
\pm 180$& $3990 \pm 277$  \\
\hline
\end{tabular}  
\caption{Glueball masses [MeV] from lattice calculations by MP
~\cite{Morningstar:1999rf}, YC~\cite{Chen:2005mg} and LTW 
~\cite{Lucini:2004my} .}
\label{Gmasses}
\end{center}
\end{table}

\begin{table} [htb]
\begin{center}
\begin{tabular} {|c c c c c c c c c|}
\hline
Meson& $f_0(500)$ &$f_0(980)$&$f_0(1370)$&$f_0(1500)$&$f_0(1710)$&$f_0(2020)$&$f_0(2100)$&$f_0(2200)$\\
\hline
PDG & $475 \pm 75$ & $990 \pm 20$ & $1350 \pm 150 $&$1504 \pm 6 $  & $1723 \pm 6 $&  $1992 \pm 16 $&  $2101 \pm 7 $&$ 2189 \pm 13$\\
\hline
\end{tabular}  
\caption{Scalar meson masses [MeV]  from PDG~\cite{Patrignani:2016xqp}}
\label{Mmasses}
\end{center}
\end{table}

\begin{figure}[htb]
\begin{center}
\includegraphics[scale= 0.75]{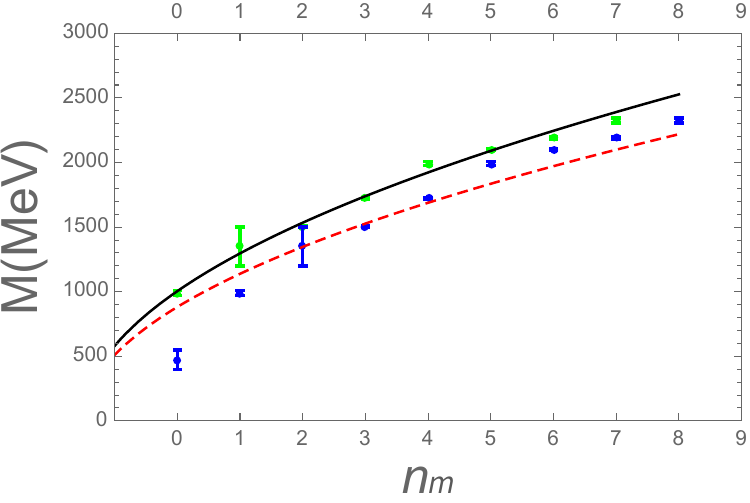} 
\end{center}
\caption{We plot the $f_0$ PDG meson spectrum~\cite{Patrignani:2016xqp}
as a function of mode number. The lower curve considers the $f_0(500)$ 
as  the lowest mass meson, while the upper curve omits this resonance 
and considers the $f_0(980)$ as the lowest mass meson. These curves 
arise by choosing the adequate energy scale in the meson spectrum of  
the soft wall model~\cite{Colangelo:2008us}.} 
\label{MesonFit}
\end{figure}

In Fig.~\ref{MesonFit} we show the soft wall model fit to the PDG meson
spectrum which we recall in Table~\ref{Mmasses}. Many authors have argued that the 
$f_0(500)$ is not a conventional meson state but a tetraquark or a 
hybrid~\cite{Mathieu:2008me,Tanabashi:2018oca}. The figure shows that 
once the $f_0(500)$ is taken out of the meson spectrum the soft wall 
model fit is excellent for $\varepsilon= 410$ MeV. From now on we omit the $f_0(500)$ from our 
discussion. 

Note the similarity of the scales for mesons and glueballs in the soft wall graviton model.
On the contrary in the soft wall model studied despite having extremely similar
 equations of motion the mass scales are very different. This feature has to do with the shape
 of the spectra. While the meson spectrum is quadratic in the region under scrutiny, the glueball
 spectrum is almost linear. 
 From this feature one may conclude that the 
soft wall graviton model describes well the 
glueball spectra. 

It might surprise that the scale factors for glueballs and  mesons are different. 
Note that  we are not approximating the same theory for mesons and glueballs. 
Fitting the mesons with PDG data we are approximating QCD, while fitting the glueballs with  
lattice QCD results we are approximating Gluodynamics. However, we feel that these scales 
should not be vastly different, an additional reason for our liking of the soft wall graviton model.
Recall that this model leads to an energy scale for glueballs of $370$ MeV and for mesons of $410$ MeV, 
which for the mass scales involved are very similar despite the different dynamics.
Lastly, remember that our fits are $\mathcal{O}(N_C^{-1})$ and higher orders should be added to 
obtain a precise value. However, the fact that the fits are quite good suggest that the 
contribution of the higher order terms might be small.
 
\begin{figure}[htb]
\begin{center}
\includegraphics[scale= 0.65]{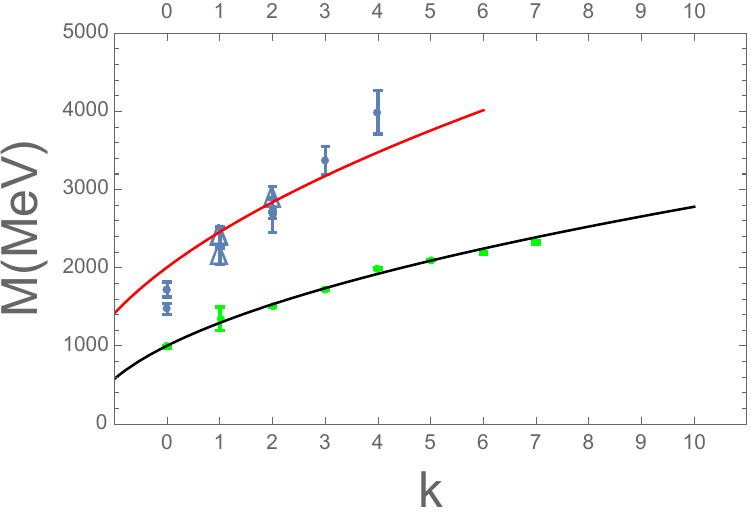}  \hspace{1cm}
\includegraphics[scale= 0.65]{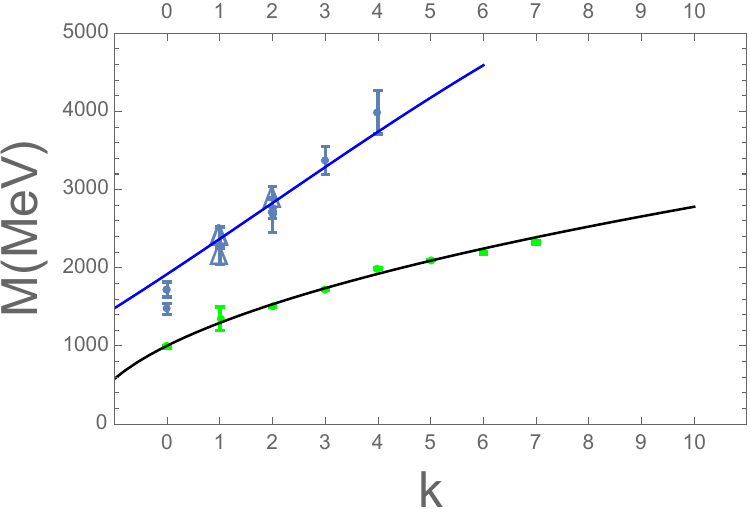} 
\end{center}
\caption{Left panel: fits of glueball spectrum (full upper line) and scalar 
meson spectrum (full lower line) obtained within the dilaton soft wall model of 
refs.~\cite{Colangelo:2007pt,Colangelo:2008us}. The dark dots represent 
glueball 
spectrum obtained within lattice QCD~\cite{Morningstar:1999rf,Chen:2005mg,Lucini:2004my}. The light dots 
represent the scalar meson spectrum obtained from experimental data ~\cite{Patrignani:2016xqp}. Right Panel: the same as in the left panel but 
the fits result from the soft wall graviton model of ref.~\cite{Rinaldi:2017wdn}.
}
\label{SpectrumFit}
\end{figure}

In Fig.~\ref{SpectrumFit} we show  the 
meson data (lower points)  and the 
glueball lattice data (upper points). We have used to fit  the meson data
 the soft wall model~\cite{Rinaldi:2017wdn,Colangelo:2008us}. To fit the data for the glueballs we have used on the left  the soft wall model 
~\cite{Colangelo:2007pt} and  on the right figure the soft wall graviton model~\cite{Rinaldi:2017wdn}. 
 An interesting feature of the comparison of spectra that can be seen  
 in Fig.\ref{SpectrumFit} is that the glueball masses with a certain mode number are equal to the meson masses with a larger mode number. 
 For example, the glueball masses for $n_g =0 , 1, 2$ are similar to the scalar meson masses for $n_m = 4, 7, 10$ respectively. The difference in mode numbers grows as the masses of the glueballs increase due to the different slopes.
  Thus besides the reasonable quality of the fits, the result we would 
like to stress is the  difference between the slopes of the glueball and 
meson fits for large mode numbers. The mode numbers are associated with the behavior of the mode functions.
Thus, the mode  function for a meson will oscillate more than that for a glueball of approximately the same mass. 
Therefore, despite having a similar mass the large difference in mode numbers leads one 
intuitively to expect that mixing will not  be very strong between the heavy states. 
We proceed to analyze the consequences of this observation 
quantitatively in what follows.

\section{Glueball-Meson mixing}

One of the problems in glueball phyics is the fact that glueball 
candidates always appear strongly mixed with mesons states~\cite{Mathieu:2008me,Vento:2015yja}.
Mixing usually occurs if two states have similar masses and the same
quantum numbers. Thus the scalar glueballs might mix with scalar 
mesons. The study of 
the $f_0$ spectrum from this perspective has led to the result that
if glueballs exist, and there is no reason for its non-existence, 
either the $f_0(1500)$ or the $f_0(1710)$
might have a large glueball component 
(see~\cite{Mathieu:2008me,Vento:2015yja} and references therein).  Our 
aim here is not to contribute to this discussion, but in view of the 
structure of the spectra,
to look for dynamical regions were mixing is not favorable and 
therefore states with mostly gluonic valence structure might exist. The 
presence of almost pure glueball states and the study of
their decays would help in understanding many properties of QCD related to
the physics of gluons.

In order to proceed with the discussion let us consider the 
holographic light-front 
representation of the equation of
motion, in $AdS$ space. The latter can be recast in the form 
of a light-front 
Hamiltonian~\cite{Brodsky:2003px}

\begin{equation}
H_{LC} |\Psi_k> = M_k^2 |\Psi_k>.
\end{equation}
where $k$ represents the mode number of the corresponding particle.
In the AdS/QCD light-front framework the above relation becomes 
a Schr\"odinger type equation

\begin{equation}
 \left(-\frac{d^2}{d z^2} + V(z) \right) \Psi(z) = M_k^2 \Psi(z)
 \label{Sch}
 \end{equation}
 where $z$ and $M_k^2$ in this equation are adimensional.
 The holographic light-front wave function are defined by $\Psi_k(z) 
 =\; <z|\Psi_k>$ and are normalized
as

\begin{equation}
<\Psi_k|\Psi_k> = \int dz |\Psi_k(z)|^2 = 1
\label{prob}
\end{equation}
The eigenmodes of Eq.(\ref{Sch}) determine the mass spectrum.  In order to fit the spectrum dimensional constants have to be introduced, e.g. the $\varepsilon$'s of previous section, which might be different for glueballs and mesons. These dimensional constants do not affect the adimensional variables and therefore do not affect the properties of the properly normalized wave functions. Thus the mode functions describe properties like probability distributions  independently of those scales in terms of the adimensional variables as seen in Eq.(\ref{prob}).

Let us discuss mixing in a two dimensional Hilbert space generated by  a meson 
and a glueball states, \{$|\Psi_m>, |\Phi_g>$\}. Mixing occurs when the hamiltonian is not 
diagonal in the subspace. Let us recall the 
discussion of two state mixing. For notational simplicity we use a 
linear hamiltonian model. A matrix representation of the hamiltonian in 
a two dimensional meson-glueball subspace is given by

\begin{equation}
[H]=  \left( \begin{array}{cc}
m_1 &  \beta  \\
\beta & m_2 \end{array} \right) ,
\label{mixing}
\end{equation}
where $\beta = <\Psi_m|H|\Phi_g>$, $m_1 = <\Psi_m|H|\Psi_m>$ 
and $m_2 = <\Phi_g|H|\Phi_g>$. We are assuming $m_2>m_1$ and for 
simplicity $\alpha$ real and positive. Large $N_C$ QCD tells us that $m_1,m_2 \sim \mathcal{O}(N_C^0)$ and $\beta \sim ~\mathcal{O}(N_C^{-1/2})$~\cite{Goity:2016ty,Vento:2004xx}. After diagonalization the eigenstates have a mass

\begin{equation}
M_{\pm}= m  \pm \sqrt{ \beta^2 + 
(\Delta m)^2},
\end{equation}
where $m=(m_1+m_2)/2$ and $\Delta m = (m_2-m_1)/2$ and its wave functions are proportional to

\[ \sim\left(
\begin{array}{c}
M_{\pm} -m_2\\
\beta\\
\end{array}
\right)\]

Thus in the starting base the physical meson, which we assume to be the lightest of the two eigenvectors, has a wave function given by

\begin{equation}
|\Psi_{phy}> =\frac{1}{\sqrt{\beta^2+ (M_- - m_2)^2}}\;((M_- - m_2)|\Psi_m> + \beta |\Phi_g>).
\end{equation}

In our fit we have fixed the meson spectrum to the experimental values and therefore $|\Psi_{phy}>$ represents a physical meson state. On the other hand we have fixed the glueball spectrum to the lattice values, i.e., to the spectrum of pure gluodynamics, therefore the glueball state is our initial state $ |\Phi_g>$, thus

\begin{equation}
|<\Psi_{phy}|\Phi_g>|^2 = \frac{\beta^2}{(M_- - m_2)^2 +\beta^2}.
\end{equation}
We conclude that the mixing probability is proportional to the overlap probability of these two wave functions.

Despite the large $N_c$ analysis previously presented one might suspect that for $N_c =3$, the overlap probability might not be as small as required to produce small $\beta$ values. The value of $\beta$ certainly depends on the modes it connects. For the lower modes it has been estimated in the linearized form of Eq.(\ref{mixing}) to be $\beta \sim 40$ MeV \cite{Giacosa:2004ug}, while the meson and glueball masses are at the level of $\sim 2000$ MeV.  The heavier meson-glueball states have smaller overlap factors, as we next show, and larger masses.  Therefore the $\beta$ values will be even smaller.

In the case of the standard soft wall dilaton model, the overlap factor is defined as

\begin{figure}[htb]
\begin{center}
\includegraphics[scale= 0.65]{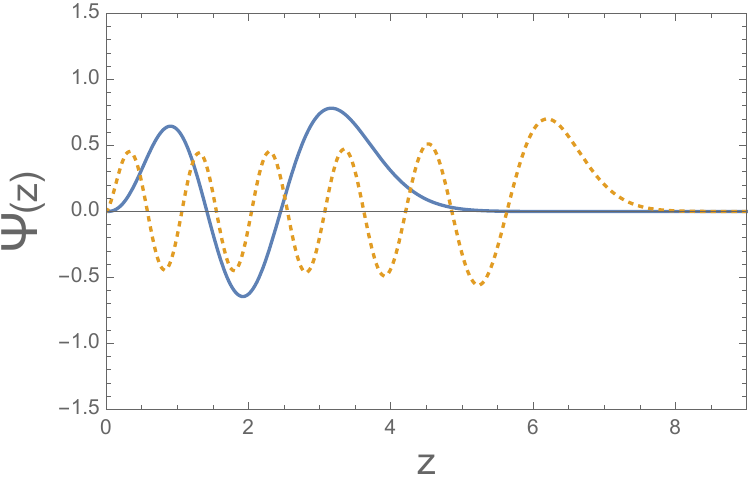} \hspace{1cm} 
\includegraphics[scale= 0.65]{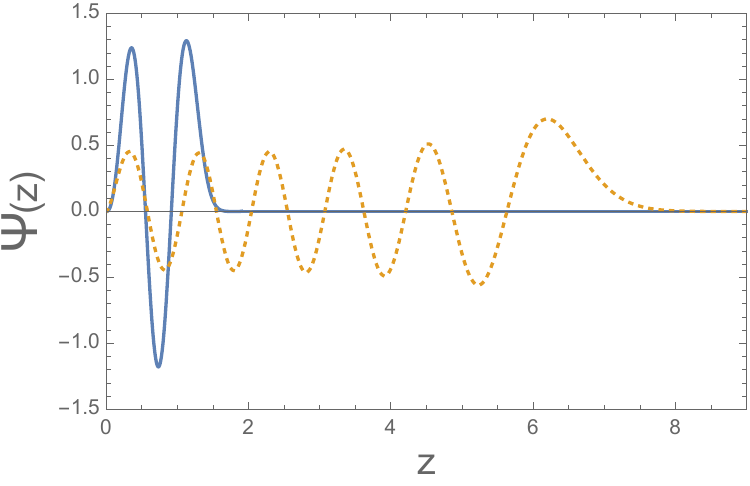} 
\end{center}
\caption{ We plot the glueball for mode number $n_g = 2$
(solid) and  the meson for mode number $n=10$ (dashed). The  figure on 
the left corresponds to the soft wall dilaton model while the figure on 
the right to the soft wall graviton model.}
\label{MesonGlueballModes}
\end{figure}

\begin{equation}
<\Psi_{n_m}|\Phi_{n_g}> = <G|M> =  \sqrt{2(n_m+1)(n_g+1)(n_g+2)}   
\int_0^\infty dz e^{-z^2} z^4 \; _1F_1(-n_g, 3, z^2) \; _1F_1(-n_m, 2, 
z^2),
\end{equation}
and the overlap probability for no mixing $P_{GM} =1- |<G|M>|^2$.
For the  soft wall graviton model the overlap factor must be obtained numerically.

\begin{figure}[htb]
\includegraphics[scale= 0.22]{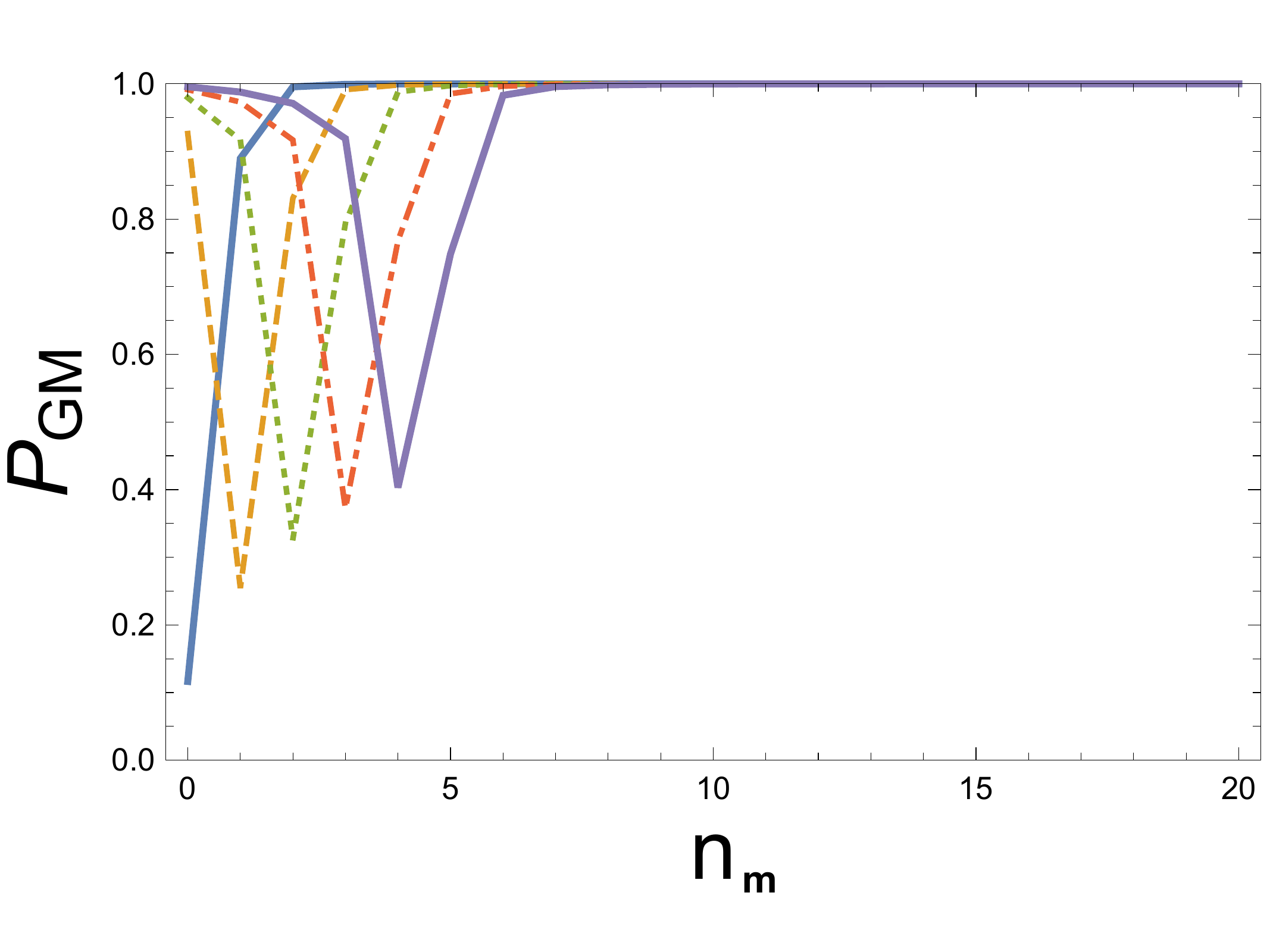} \hspace{1.0cm}
\includegraphics[scale= 0.22]{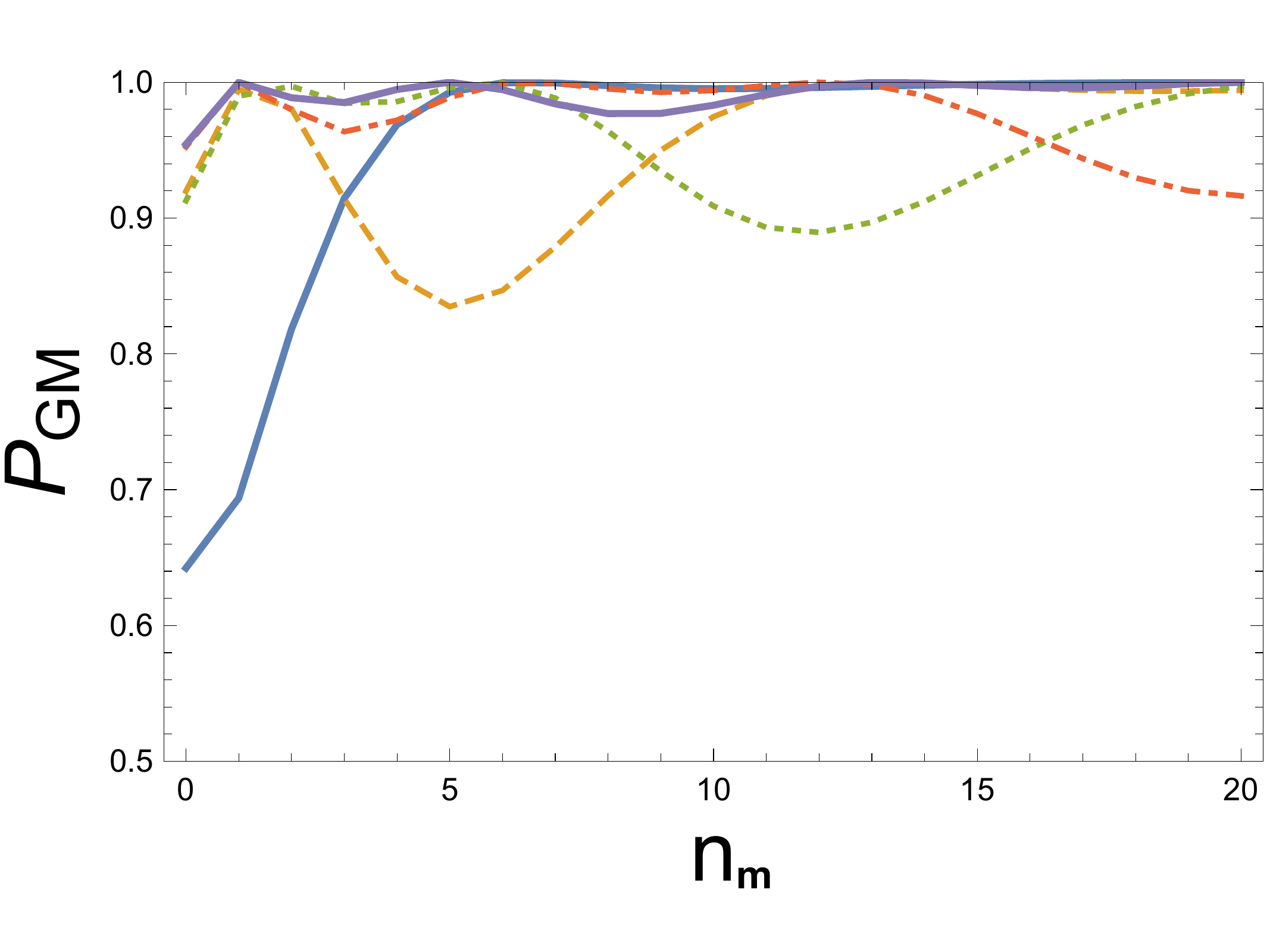} 
\caption{ We plot the probability of no mixing for the glueball with
mode numbers $n_g = 0$ (solid),$1$ (dashed), $2$ (dotted), $3$ 
(dot-dashed), $4$ (solid) as a function of meson mode number $n_m$. The 
left figure is for the soft wall dilaton model while the right figure is 
for the soft wall graviton model.}
\label{OverlapDilatonGraviton}
\end{figure}

We would like to show that meson and glueball wave functions whose mode 
numbers are very different, despite having almost equal masses, lead to 
small overlap probabilities and therefore to small mixing. 
To do so let us an example. We choose a glueball of mode number $n_g=2$
and a meson of mode number $n_m=10$. They  can be 
considered as candidates for  mixing because, using in the soft wall model Eqs.~(\ref{glueballsp},\ref{mesonsp}),  the glueball  mass $m(k=2) \sim 2840$ 
MeV  and the meson mass $m(k=10) \sim 2781 $ 
MeV. Thus this example can be considered as a prototype for a mixing scenario 
for heavy particles.   In Fig.\ref{MesonGlueballModes} we show the mode functions 
for the $n_g=2$ glueball mode and that for the $n_m =10$ meson mode for both the soft wall and soft wall graviton models. 
The figure shows  the big difference 
between the glueball modes in both the soft wall and soft wall graviton models. The extension of the glueball 
mode in the soft wall graviton model  is determined by the exponential 
in the potential and therefore all modes die at $z \sim 2$ irrespective 
of their mode number. In the case of the soft wall  model the 
extension is governed by the mode number. This difference in structure 
implies that in the soft wall graviton model the overlap factor is oscillating and 
extends over many modes even reaching  large meson mode numbers $n$, 
while in the dilaton model only a few modes close to $n_g$ are 
contributing to the overlap as shown in Fig. 
\ref{OverlapDilatonGraviton}.

Looking at Figs.~\ref{SpectrumFit} and \ref{OverlapDilatonGraviton},
the favorable mixing scenario is mostly excluded in the case of heavy 
glueballs and mesons, since the  mass condition is satisfied for very 
different mode numbers. For example $n_g=2,3,4$ the favorable meson modes 
of almost equal masses occur 
for  $n_m \sim 10,13,17$ in the soft wall model~\footnote{In the soft wall graviton model the difference is even larger 
since the slope of the glueball fit is larger than that of the soft wall model.}.
As can be seen in Fig.~\ref{OverlapDilatonGraviton} this condition reduces the  overlap probability for 
mixing dramatically. In the soft wall  model the overlap 
probability is extremely small for the required mode number differences, 
while in the soft wall graviton model it oscillates at the level of maximum  
$10\%$ percent overlap probabilities. The outcome of our analysis is 
that the AdS/QCD approach predicts the existence of almost pure 
glueball states in the scalar sector in the mass range above $2$ GeV.
The no mixing  scenario is experimentally very appealing. It predicts the existence of pairs of almost degenerate 
states with very different decay properties. The pure glueball decays to mesons are Zweig 
forbidden, i.e.  small widths and an increase in the proportion of strange mesons to non-strange mesons \cite{Mathieu:2008me,Klempt:2007cp,Crede:2008vw,Ochs:2013gi}. 
On the contrary the pure meson states will have a large width since they have a lot of phase space for decaying and larger decay rates to non-strange mesons than to strange mesons.

\section{Conclusion}
We have performed a phenomenological analysis of the scalar glueball
and scalar
meson spectrum based on the  AdS/QCD 
correspondence within the soft wall dilaton and soft wall graviton approaches.
Theoretical outcomes have been compared with lattice $QCD$ data in the case of the 
glueballs and the experimental $f_0$ spectrum of the PDG tables in the case of the mesons.   We have 
noted that the slope of the glueball spectrum as a function of mode 
number is bigger that that of the meson spectrum in both approaches
and therefore for heavy almost degenerate glueball and meson 
states, their mode numbers differ considerably. Assuming a light-front 
quantum mechanical description of AdS/QCD correspondence we have shown 
that the overlap probability of heavy glueballs to heavy mesons is small and 
thus one expects little mixing in the high mass sector. Therefore, 
this is the kinematical region to look for almost pure glueball states. The scenario is 
phenomenologically very appealing because it implies the existence of pairs of almost degenerate 
states with very different decay properties.

\section*{Acknowledgments}
We acknowledge Risto Orava, Sergio Scopetta, Tatiana Tarutina  and Marco Traini for
discussions. VV thanks the hospitality extended to him by the 
University of Perugia  and the INFN group in Perugia. MR was a Severo 
Ochoa postdoctoral fellow at IFIC
during the initial stages of this work.This work was supported  
by MICINN and UE Feder under contract FPA2016-77177-C2-1-P, by Severo Ochoa Program under contract
SEV-2014-0398 and by the STRONG-2020 project of the European UnionÕs Horizon 2020 research and innovation
programme under grant agreement No 824093.

\end{document}